\documentclass[preprint,11pt]{elsarticle}

\usepackage{graphicx}
\usepackage{amsmath}
\usepackage{amssymb}
\usepackage{hyperref}
\usepackage{dcolumn}
\usepackage{bm}
\usepackage{color}

\begin{document}

\begin{frontmatter}

\title{Invariant currents and scattering off locally symmetric potential landscapes}

\author{P.~A.~Kalozoumis}


\address{Zentrum f\"ur Optische Quantentechnologien, Universit\"{a}t Hamburg, Luruper Chaussee 149, 22761 Hamburg, Germany}

\address{Department of Physics, University of Athens, GR-15771 Athens, Greece}

\author{C.~Morfonios}


\address{Zentrum f\"ur Optische Quantentechnologien, Universit\"{a}t Hamburg, Luruper Chaussee 149, 22761 Hamburg, Germany}

\author{F.~K.~Diakonos}


\address{Department of Physics, University of Athens, GR-15771 Athens, Greece}

\author{P.~Schmelcher}


\address{Zentrum f\"ur Optische Quantentechnologien, Universit\"{a}t Hamburg, Luruper Chaussee 149, 22761 Hamburg, Germany}

\address{Hamburg Centre for Ultrafast Imaging, Universit\"{a}t Hamburg,
Luruper Chaussee 149, 22761 Hamburg, Germany}

\date{\today}

\begin{abstract}

We study the effect of discrete symmetry breaking in inhomogeneous scattering media within the framework of generic wave propagation. Our focus is on one-dimensional scattering potentials exhibiting local symmetries. We find a class of spatially invariant nonlocal currents, emerging when the corresponding generalized potential exhibits symmetries in arbitrary spatial domains. These invariants characterize the wave propagation and provide a spatial mapping of the wave function between any symmetry related domains. This generalizes the Bloch and parity theorems for broken reflection and translational symmetries, respectively. Their nonvanishing values indicate the symmetry breaking, whereas a zero value denotes the restoration of the global symmetry where the well-known forms of the two theorems are recovered. These invariants allow for a systematic treatment of systems with any local symmetry combination, providing a tool for the investigation of the scattering properties of aperiodic but locally symmetric systems. To this aim we express the transfer matrix of a locally symmetric potential unit via the corresponding invariants and derive quantities characterizing the complete scattering device which serve as key elements for the investigation of transmission spectra and particularly of perfect transmission resonances. 

\end{abstract}


\end{frontmatter}

\section{Inroduction}  \label{intro}

Symmetries constitute one of the cornerstones of physics, being prominently displayed due to their fundamental role in the theoretical treatment of any system. Symmetry principles not only reduce the extent of information which is required for the description of a physical system but also dictate the form of physical laws. Under this prism, the endeavour to extend the frontiers of our knowledge about nature is significantly based on the discovery of higher symmetry principles.

The usual pathway is to consider symmetry principles which hold \textit{globally} for a physical system, explaining important phenomenological properties and facilitating the mathematical description.
However, global symmetry usually is an idealized scenario, mainly met in models, approximative schemes or structurally simple isolated systems. The concept of local symmetry is usually introduced in the context of gauge transformations involving space-time dependent parameters which, in turn, imply that the associated symmetry is valid at a single space-time point.

Between these two symmetry classes, the one valid at every point of space (or space-time) and the other valid at a single point, another category can be defined, where different symmetries are fulfilled in different spatial domains of finite extent. Physical systems possessing the latter property usually emerge due to the breaking of a global symmetry so that new symmetries at different scales occur. Such symmetry-associated patterns are manifest in extensively diverse structures encountered in nature~\cite{echeverria2011} and dominate several length scales. Therefore, it is a generic situation to deal with extended physical systems involving domains which are locally characterized by a certain symmetry.

Spatially localized symmetries can be intrinsic in complex systems such as, e.g., large molecules~\cite{Grzeskowiak1993,Pascal2001,Chen2012}, quasicrystals~\cite{Shechtman1984,Levine1984,Widom1989,Lifshitz1996}, self-organized, pattern-forming systems~\cite{pattern} or partially disordered matter~\cite{Wochner2009}.  On the other hand, they can be present by design in multilayered photonic devices~\cite{Macia2006,Zhukovsky2010,Peng2002,Peng2004,Hsueh2011,Vardeny2013}, quantum semiconductor superlattices~\cite{Ferry1997}, acoustic waveguides~\cite{Hladky2013} or magnonic systems~\cite{Hsueh2013}. Moreover,  technological advances often require the breaking of global discrete symmetries in order to obtain flexible structures suitable for applications. 

Despite the fact that systems belonging to the aforementioned classes have been extensively investigated, the theoretical framework for their mathematical description is usually restricted to the case of global symmetries.
On a local level, studies on the structural features which affect  spectral and localization properties  has been carried out~\cite{Macia2012}, for instance in hybrid systems which are comprised of domains each with different quasiperiodic structure. However,  little attention has been paid to the impact of explicit {\it local  symmetries}, implied by the breaking of a global symmetry, although they are obviously present and often coexist at different spatial scales. A rigorous theoretical treatment which addresses the local symmetry induced properties is missing and obviously a new point of view needs to be introduced.

A step towards this direction was recently made by introducing the concept of \textit{local parity} (LP) \cite{Kalozoumis2013a}. It was shown how a system's decomposition into  mirror symmetric units relates to spectral properties and especially to perfect transmission resonances (PTRs) in aperiodic setups. The origin of perfect transmission in such systems, even though tentatively linked to symmetry concepts~\cite{Huang2001,Nava2009}, had yet been an unresolved issue, lacking a rigorous theory which directly relates perfect transmission  in aperiodic and quasiperiodic setups to  its underlying (local) symmetry properties. 
Within the local parity approach, an unambiguous symmetry based classification of scattering states has been established \cite{Kalozoumis2013b}, elucidating the link between perfect transmission and spatial domain symmetry. 

A classification of scattering states based on a setup's decomposition in locally symmetric domains leads to the question whether the identification of remnants of the broken global symmetry is possible and if so, how these remnants may determine the form of the scattered wave in the symmetry related domains in the same manner the Bloch~\cite{Bloch1929} and parity~\cite{Zettili2009} theorems do when the corresponding translation and reflection symmetry are globally satisfied (we use here the term {\it parity theorem} for the well known theorem of quantum mechanics concerning the commutation of the global inversion operator with the Hamiltonian and the existence of common eigenstates for both operators).  Subsequently, another question is natural to be posed: What form, if any, will the generalized parity and Bloch theorems acquire if global reflection or translation symmetry is broken? The answer was given recently in Ref.~\cite{Kalozoumis2014a}, where a systematic pathway to the symmetry breaking of discrete symmetries was derived, generalizing the Bloch and parity theorems for broken translational and reflection symmetries, respectively. This formalism can handle connected or disconnected arbitrary, symmetric spatial domains of finite or infinite extent.

The aim of the present work is twofold. Firstly, we extend the theoretical framework established in~\cite{Kalozoumis2014a}, emphasizing on the derivation of new properties of the scattered waves, which emerge from their phase and magnitude. The analysis is based on the invariant non-local currents induced by symmetry breaking. Secondly, we introduce sum rules involving these invariants and we demonstrate how they are related to the transmission properties of 
waves which propagate in devices which are completely decomposable into locally symmetric subunits. In order to be self-contained we include in our presentation the derivation of the nonlocal currents and we demonstrate how they allow for the generalization of the Bloch and parity theorems for arbitrary setups. 

The paper is organized as follows: In Sec.~\ref{invariants} we derive a pair of symmetry induced nonlocal invariant currents for the discrete reflection and translation symmetry operations, which exist in arbitrary symmetry domains of a generic wave scattering setup.  In Sec.~\ref{Bloch} we show how these currents allow the mapping of the wave function between the symmetry related domains, generalizing the Bloch and parity theorems. In the limit of the global symmetry restoration we recover the usual forms of the theorems. In Sec.~\ref{scattering} we demonstrate how scattering from aperiodic systems can be alternatively formulated by expressing the transfer matrix of a locally symmetric potential unit via the corresponding invariants and in Sec.~\ref{scales} we comment on the significance of the possible decompositions of a setup in multiple symmetry scales. Using sum rules of the corresponding invariants we derive quantities which describe globally aperiodic devices which can be decomposed in locally symmetric domains and investigate their energy dependence and their relation to the transmission spectrum. Our conclusions are provided in Sec.~\ref{conclusions}.

\section{Discrete symmetry induced invariant nonlocal currents} \label{invariants} 

\begin{figure}[t!]
\centering
\includegraphics[width=.85\columnwidth]{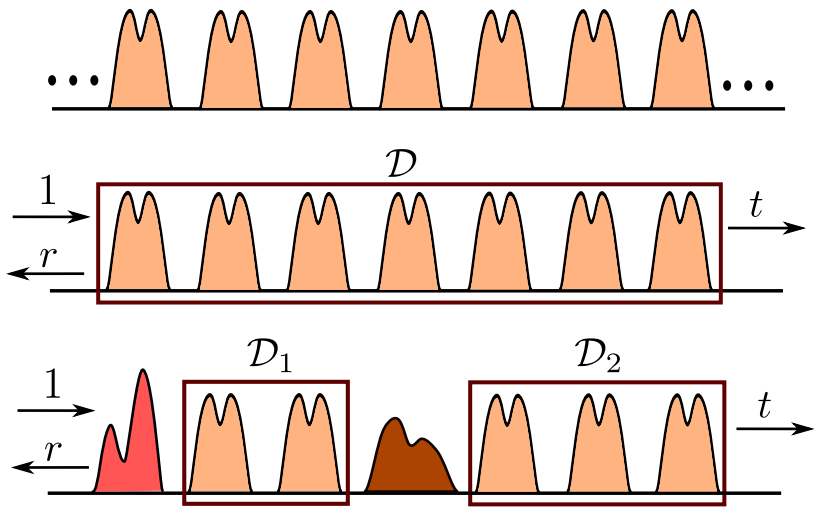}
\caption{\label{fig1} (Color online) (a) Schematic illustration of a periodic system with global translational symmetry (extending to infinity). (b) Finite periodic system. Even though the device or potential shows a discrete translational symmetry on its domain of definition, the symmetry is in particular broken due to the asymmetric asymptotic conditions.
(c) The addition of two defects explicitly breaks the local translational symmetry within $\mathcal{D}$, which now is retained locally in the restricted spatial domains $\mathcal{D}_{1},~\mathcal{D}_{2}$. }
\end{figure}

We first link the concept of translational and reflection symmetry to the properties of stationary waves described by the Helmholtz equation.
A globally symmetric system is one which exhibits the corresponding symmetry for all $x \in \mathbb{R}$. An infinite, periodic system as the  one illustrated in Fig.~\ref{fig1}(a) belongs to this class. However, the generic case in realistic physical systems, is that global symmetries are broken. Two cases of this symmetry breaking can be identified:

(i) The asymptotic conditions violate the symmetry whereas a central part of the setup still preserves some symmetry. This is e.g. the case for a scattering situation with asymmetric asymptotic waves while the scattering potential still retains the corresponding symmetry property for all $x \in \mathcal{D}$, where $\mathcal{D}$ is the scattering region where the potential acts (see Fig.~\ref{fig1}(b)). 

(ii) The potential does not exhibit the corresponding symmetry $\forall x \in \mathcal{D}$.
In the latter case the symmetry can be either completely absent or maintained \textit{locally} in spatially restricted domains. This situation is illustrated in Fig.~\ref{fig1}(c), where the translational symmetry of potential in Fig.~\ref{fig1}(b) is broken due the presence of two defects. Nevertheless, the symmetry is retained on a local level within the domains $\mathcal{D}_{1},~\mathcal{D}_{2}$, showing how a broken symmetry can give rise to local symmetries (local parity or local translation) valid in restricted spatial domains. Of special interest is the case of systems which can be completely decomposed into parts that fulfill certain reflection or translation symmetries exactly. Such systems form the special class of {\it completely locally symmetric (CLS)} systems  which extend the notion of periodicity or global parity symmetry.  In the following sections we will extensively study such systems, revealing their intriguing properties. 

Let us consider such a generic wave scattering system with generalized potential $U(x)=\kappa^{2}(x)$. The generalized wave vector $\kappa(x)$ describes the inhomogeneities of the medium where the wave propagation occurs. Here we will consider the system to be homogeneous in the $yz$-plane and  varying only in the $x$-direction.  We also restrict the wave to normal incidence on the $yz$-plane, such that it propagates  along the $x$-axis. Then the wave field can  be written $\mathcal{A}(\vec{x},t) = \mathcal{A}(x)e^{-i\omega t}\hat{x}$, where $\mathcal{A}(x)$ is the complex field amplitude. Using this notation, the developed formalism can be applied directly not only to quantum mechanical but also to optical, acoustic or other wave mechanical system described by the Helmholtz equation. 
For instance, in the quantum mechanical case, $\mathcal{A}(x)$ represents the wave function, with $U(x)$ would be $U(x)=(2 m/ \hbar^2)[\varepsilon-V(x)]$, with $m$ being the mass and $\varepsilon$ the energy of the quantum particle in the potential $V(x)$. For electromagnetic waves with frequency $\omega$, the function $\mathcal{A}(x)$ could represent the electric field, while $U(x)=\omega^2 n^2(x)/c^2$ ($n(x)$ being the refractive index of the medium of propagation).

\subsection{Invariant nonlocal currents} 

\noindent Focusing on systems for which $U(x)$ exhibits local symmetries we proceed by showing first the existence of nonlocal currents which are spatially constant within the finite domain(s) where the medium obeys a specific symmetry. Subsequently we demonstrate how these currents can be used to determine the structure of the solution of the associated wave equation in the symmetry related domain(s), generalizing the Bloch and parity theorems for systems with reflection or translational symmetry in restricted spatial domains. The analysis and the presentation of our theory is performed in one dimension assuming the wave propagation to be described by the Helmholtz equation. However, it can be extended to higher dimensions in which case the translation symmetry can be directly generalized while the reflection symmetry is replaced by the inversion with respect to a point or reflection through a plane. 

In order to treat wave propagation in inhomogeneous media within a unified framework   we employ the  Helmholtz equation:
\begin{equation}
\label{helmholtz} \mathcal{A}''(x)+U(x)\mathcal{A}(x)=0.
\end{equation}
where the prime denotes differentiation with respect to $x$.
We consider the following linear transform
\begin{equation}
\label{F}
F(x)\equiv \bar{x}=\sigma x + \rho~;~ \left\lbrace \begin{array}{l} \sigma=-1~;~\rho=2 \alpha~(\mathrm{reflection}) \\ \sigma=+1~;~\rho=L~~(\mathrm{translation}) \end{array} \right.
\end{equation}
which acts on the generalized potential of Eq.~(\ref{helmholtz})  in the following manner:
\begin{equation}
\label{potinv}
F(U(x))=U(\bar{x}),
\end{equation}
$\forall~x$ in the corresponding symmetric domain.
The transform in Eq.~(\ref{F}) describes a reflection about the point $\alpha$ when 
$\sigma=-1$ and a translation by $L$ when $\sigma=+1$. 

Since Eq.~(\ref{helmholtz}) is valid for every $x$ in $\mathbb{R}$ it must also hold for the image of $x$ under the transform $F$:
\begin{equation}
\label{h_transf} \mathcal{A}''(\bar{x})+U(\bar{x})\mathcal{A}(\bar{x})=0.
\end{equation}
Now we multiply Eq.~(\ref{helmholtz}) by $\mathcal{A}(\bar{x})$ and Eq.~(\ref{h_transf}) by $\mathcal{A}(x)$. Subsequently, we subtract the resulting equations from each other,  taking into account the symmetry of the generalized potential $U(x)$ (valid only for $x \in \mathcal{D}$), expressed by Eq.~(\ref{potinv}).  As a result we obtain:
\begin{equation}
\label{conserv1}
\mathcal{A}(\bar{x})\mathcal{A}''(x)-\mathcal{A}(x) \mathcal{A}''(\bar{x})=0
\end{equation}
Equation~(\ref{conserv1}) (for $\sigma=\pm 1$) has the form of a total derivative:
\begin{equation}
\label{totder}
\frac{d}{dx} \left[ \mathcal{A}(\bar{x}) \mathcal{A}'(x) -\sigma \mathcal{A}(x) \mathcal{A}'(\bar{x})\right]=0~~~; ~~~\forall~~x \in \mathcal{D}
\end{equation}
which in turn implies that the complex quantity:
\begin{equation}
\label{Q}
Q=\frac{1}{2i}\left[ \sigma \mathcal{A}(x)
\mathcal{A}'(\bar{x})- \mathcal{A}(\bar{x}) \mathcal{A}'(x) \right]
\end{equation}
is spatially invariant within the domain $\mathcal{D}$, where the respective symmetry is fulfilled. 

In the same manner, we can use the complex conjugate of Eq.~(\ref{helmholtz}) (or Eq.~(\ref{h_transf})) and repeat the same procedure. Then we obtain another independent, spatially invariant quantity in the domain 
$\mathcal{D}$:
\begin{equation}
\label{Qtilde}
\widetilde{Q}=\frac{1}{2i}\left[\sigma \mathcal{A}^{*}(x) \mathcal{A}^{\prime}(\bar{x})-\mathcal{A}(\bar{x}) \mathcal{A}^{\prime *}(x)\right] 
\end{equation}
The invariant quantities defined by Eqs.~(\ref{Q}),~(\ref{Qtilde}) have the form of a {\it nonlocal current}, involving points connected by the corresponding symmetry transform. Since we refer to a real generalized potential $U(x)$ in Eq.~(\ref{helmholtz}), apart from the constants $Q$ and $\widetilde{Q}$, there exists also the globally conserved local current $J$ given by:
\begin{equation}
\label{J}
J=\frac{1}{2i}\left[\mathcal{A}^{\prime}(x)\mathcal{A}^{*}(x)-\mathcal{A}^{\prime *}(x)\mathcal{A}(x)\right]
\end{equation}
which represents the probability current in the quantum mechanical case or the $1$D analogue of the Poynting vector~\cite{Chen1987} in the electromagnetic case. The invariants $Q$, $\widetilde{Q}$, $J$ are linked via the relation
\begin{equation}
\label{QJ}
\sigma \left(\vert \widetilde{Q} \vert^2 - \vert Q \vert^2 \right) = J^2.
\end{equation}
Equation~(\ref{QJ}) can be directly obtained by subtracting the moduli of $Q,~\widetilde{Q}$ from Eqs.~(\ref{Q}),~(\ref{Qtilde}) and using $J(x)=J(\bar{x})$.

\section{Generalization of the Bloch and parity theorems for broken symmetries} \label{Bloch}
 
In its most general form,  the transform $F(x)=\bar{x}$ maps a domain $\mathcal{D}$  to a different domain $\overline{\mathcal{D}}$. These domains do not need to be connected; they can be separated by any distance, as long as the symmetry is preserved. In the usual case of parity, where the mirror axis at $\alpha$ belongs to the domain, the mapping occurs from $\mathcal{D}$ onto itself and particularly the domain on the right-hand side (rhs) of the mirror axis is mapped onto the domain on the left-hand side (lhs) and vice-versa.

The image of the wave field $\mathcal{A}(\bar{x})$, can be expressed in terms of 
$\mathcal{A}(x)$, $\mathcal{A}^*(x)$ and the invariants $Q$, 
$\widetilde{Q}$ by solving the system of Eqs.~(\ref{Q}),~(\ref{Qtilde})  with respect to $\mathcal{A}(\bar{x})$ and $\mathcal{A}'(\bar{x})$. If $U(x)=U(\bar{x})$,this in turn yields
\begin{equation}
\label{genblp}
\mathcal{A}(\bar{x})=\frac{\widetilde{Q}}{J} \mathcal{A}(x) -
\frac{Q}{J} \mathcal{A}^*(x)
\end{equation}
and
\begin{equation}
\label{ader}
\mathcal{A}'(\bar{x})=\sigma \left(\frac{\widetilde{Q}}{J} \mathcal{A}'(x) - \frac{Q}{J} \mathcal{A}^{\prime *}(x)\right)
\end{equation} 
Equation~(\ref{genblp}) is of central importance both to the theoretical concept of local symmetries as well as the concrete properties of locally symmetric devices. One can obtain a direct mapping of the image $\mathcal{A}(\bar{x})$ in the target domain 
$\overline{\mathcal{D}}$ from $\mathcal{A}(x)$ in $\mathcal{D}$ by using the constant nonlocal currents $Q$ and $\widetilde{Q}$ which are a result of the underlying symmetry of $U(x)$. In this sense, it constitutes the generalization of the Bloch and parity theorems in the case of a generalized potential $U(x)$ with a broken global symmetry. As will be shown in the following, a \textit{nonvanishing invariant} current $Q$ is the manifestation of the broken global symmetry.

The relation~(\ref{genblp}) allows to derive the corresponding map for the magnitude and the phase of the wave function $\mathcal{A}(x)$. In order to simplify the resulting expressions we write $Q,~\widetilde{Q}$ as:
\begin{equation}
\label{Q_via_real_q} Q=q_{1}-q_{4}+i(q_{2}+q_{3})
\end{equation}
and
\begin{equation}
\label{Qtilde_via_real_q} \widetilde{Q}=q_{1}+q_{4}+i(q_{2}-q_{3})
\end{equation}
where $q_{1},~q_{2},q_{3}$ and $q_{4}$ are \textit{imaginary} spatially invariant quantities within the corresponding symmetry domains, given by:
\begin{equation}
\label{q1_invariant} q_{1}=\frac{1}{2i}\operatorname{Re}\left[ \frac{Q+\widetilde{Q}}{2} \right]=\frac{1}{2i}\left[\sigma \operatorname{Re}[\mathcal{A}'(\bar{x})] \operatorname{Re}[\mathcal{A}(x)]-\operatorname{Re}[\mathcal{A}(\bar{x})] \operatorname{Re}[\mathcal{A}'(x)]\right],
\end{equation}
\begin{equation}
\label{q2_invariant} q_{2}=\frac{1}{2i}\operatorname{Im}\left[ \frac{Q+\widetilde{Q}}{2} \right]=\frac{1}{2i}\left[\sigma \operatorname{Im}[\mathcal{A}'(\bar{x})] \operatorname{Re}[\mathcal{A}(x)]-\operatorname{Im}[\mathcal{A}(\bar{x})] \operatorname{Re}[\mathcal{A}'(x)]\right],
\end{equation}
\begin{equation}
\label{q3_invariant} q_{3}=\frac{1}{2i}\operatorname{Re}\left[ \frac{Q-\widetilde{Q}}{2i} \right]=\frac{1}{2i}\left[\sigma \operatorname{Re}[\mathcal{A}'(\bar{x})] \operatorname{Im}[\mathcal{A}(x)]-\operatorname{Re}[\mathcal{A}(\bar{x})] \operatorname{Im}[\mathcal{A}'(x)]\right],
\end{equation}
\begin{equation}
\label{q4_invariant} q_{4}=\frac{1}{2i}\operatorname{Im}\left[ \frac{Q-\widetilde{Q}}{2i} \right]=\frac{1}{2i}\left[\sigma \operatorname{Im}[\mathcal{A}'(\bar{x})] \operatorname{Im}[\mathcal{A}(x)]-\operatorname{Im}[\mathcal{A}(\bar{x})] \operatorname{Im}[\mathcal{A}'(x)]\right].
\end{equation}
In this notation it can be shown that the current $J$ has the form:
\begin{equation}
\label{cur_q} J^{2}=q_{2}q_{3}-q_{1}q_{4}.
\end{equation} 

Using the polar representation $\mathcal{A}(x)=u(x)e^{i\varphi(x)}$ and after some extensive but straightforward algebraic manipulations, we find: 
\begin{equation}
\label{magn_map} u(\bar{x})=u(x)\left[\frac{q_{3}^{2}+q_{4}^{2}-2(q_{1}q_{3}+q_{2}q_{4})\tan[\varphi(x)] + (q_{1}^{2}+q_{2}^{2}) \tan^{2}[\varphi(x)]}{J^{2}\left(1+\tan^{2}[\varphi(x)]\right)} \right]^{1/2}
\end{equation}
and
\begin{equation}
\label{phase_map} \tan[\varphi(\bar{x})]=\frac{q_{4}-q_{2}\tan[\varphi(x)]}{q_{3}-q_{1}\tan[\varphi(x)]}.
\end{equation}
The constant quantities $q_{1},~q_{2},~q_{3},~q_{4}$ thus provide a direct mapping between the symmetry related domains for the phase $\varphi$ of the wave function. On the contrary, for the respective mapping for the magnitude $u(x)$ the information of the phase is necessary.

\subsection{Globally symmetric potentials}

\noindent In order to explain transparently the mechanism of symmetry breaking 
we define the linear operator $\hat{O}_F$ which acts in the coordinate representation on an arbitrary function $\varPhi(x)$ and transforms it according to the respective symmetry operation:
\begin{equation}
\label{oper}
\hat{O}_F \varPhi(x) = \varPhi(\bar{x})~~~~~~;~~~~~~\forall~~x~\in~\mathbb{R}.
\end{equation}
Global symmetry with respect to the transform $\hat{O}_F$ is realized when $U(x)=U(\bar{x})$ for all $x~\in~\mathbb{R}$ and the Helmholtz operator
\begin{equation}
\label{gen_helm_eq}\hat{\varOmega}=\frac{d^2}{dx^2}+U(x)
\end{equation}
commutes with $\hat{O}_F$, rendering 
$\mathcal{A}(x)$ an eigenstate of $\hat{O}_F$:
\begin{equation}
\label{eigenO}
\hat{O}_F \mathcal{A}(x)=\lambda_F \mathcal{A}(x),
\end{equation}
where $\lambda_F$ is the respective eigenvalue.
Particularly, the translation operator $\hat{T}_{L}$ which causes a translation by $L$ is defined via the relation :
\begin{equation}
\label{trans_operator} \hat{T}_{L}\mathcal{A}(x)=\mathcal{A}(x+L). 
\end{equation}
If $\mathcal{A}(x)$ is an eigenfunction of  $\hat{T}_{L}$, then we can write:
\begin{equation}
\label{trans_operator_3}\hat{T}_{L}\mathcal{A}(x)=\lambda_{T}\mathcal{A}(x).
\end{equation}
When the setup extends to infinity, the Bloch theorem determines the form of $\mathcal{A}(x)$:
\begin{equation}
 \mathcal{A}(x+L)=e^{ikL}\mathcal{A}(x),
\end{equation}
where $\hbar k$ is the crystal momentum. We can then write:
\begin{equation}
\hat{T}_{L} \mathcal{A}(x)=e^{ikL}\mathcal{A}(x)
\end{equation}
and therefore the eigenvalues of $\hat{T}_{L}$ are phases $\lambda_{T}=e^{ikL}$ lying on the unit circle. 

Similarly, the parity operator $\hat{\varPi}$ acting on $\mathcal{A}(x)$ yields:
\begin{equation}
\hat{\varPi} \mathcal{A}(x)=\mathcal{A}(-x).
\end{equation}
If $\mathcal{A}(x)$ is an eigenfunction of $\hat{\varPi}$ then:
\begin{equation}
\hat{\varPi} \mathcal{A}(x)=\lambda_{\varPi}\mathcal{A}(x)
\end{equation}
and straightforwardly:
\begin{equation}
\mathcal{A}(x)=\hat{\varPi}^{2} \mathcal{A}(x)=\lambda_{\hat{\varPi}}^{2}\mathcal{A}(x).
\end{equation}
where the parity eigenvalues are $\lambda_{\varPi}=\pm1$ correspond to even and odd eigenfunctions.

The simplest scenario to break the global $\hat{O}_F$ symmetry is when $\hat{{\varOmega}}$ still commutes with $\hat{O}_F$, i.e. $$U(x)=U(\bar{x})~~~ \forall x \in \mathbb{R}$$
but $\mathcal{A}(x)$ ceases to be an eigenfunction of $\hat{O}_F$ violating Eq.~(\ref{eigenO}) due to its asymptotic behaviour, which is typically the case in a scattering problem (see Fig.~\ref{fig1} (b)). Remarkably, within the present framework, even if the symmetry is broken due the asymptotic conditions the conserved quantities $Q$ and $\widetilde{Q}$ are constant in the entire space and Eq.~(\ref{genblp}) applies for all $x$ in $\mathbb{R}$ due to the global underlying symmetry of the potential. 
Using Eqs.~(\ref{genblp}) and (\ref{oper}) we can write:
\begin{equation}
\label{eigf}
\hat{O}_F \mathcal{A}(x) = \frac{\widetilde{Q}}{J} \mathcal{A}(x) -
\frac{Q}{J} \mathcal{A}^*(x)~~~~~;~~~~~\forall~~x~\in~\mathbb{R}
\end{equation}
which clearly shows that  $Q \neq 0$ manifests itself as a \textit{remnant} of the broken global translation or reflection symmetry. 

\subsubsection{Retrieving the Bloch and parity theorems}

\noindent To set $Q=0$ has interesting consequences on the field $\mathcal{A}(x)$. One can integrate Eq.~(\ref{Q}) and get
\begin{equation}
\label{inv}
\mathcal{A}(\bar{x})=c \mathcal{A}(x)
\end{equation}
where $c \in \mathbb{C}$ is an integration constant. If however we set $c=\lambda_F=\frac{\widetilde{Q}}{J}$ we recover Eq.~(\ref{eigenO}), which is consistent with our interpretation of the invariant $Q$ as a symmetry breaking term. Based on the vanishing of $Q$, we will show rigorously how the parity and Bloch theorems are retrieved in the limit of global symmetry restoration.
This completes the argumentation on how Eq.~(\ref{genblp}) generalizes the parity and Bloch theorems for the case of broken global symmetry. Let us discuss in the following how the parity and Bloch theorems are retrieved for the case of a global reflection and translation symmetry, respectively.

\noindent \textit{Reflection} ($\sigma=-1$): Starting from the reflection case we integrate Eq.~(\ref{Q}) assuming $Q=0$ which, as previously discussed,  is a necessary condition for a global discrete symmetry to hold. This leads to:
\begin{equation}
\label{gp}
\mathcal{A}(2 \alpha -x)= c \mathcal{A}(x)~~~;~~~\mathcal{A}'(2 \alpha -x)= -c \mathcal{A}'(x)
\end{equation}
where $c$ is an integration constant. One can determine $c$ by setting in Eq.~(\ref{Q}) $x=\alpha$ since for the case of global parity the symmetry axis necessarily belongs to the domain of mirror symmetry which is the entire space. This leads to $\mathcal{A}(\alpha)\mathcal{A}^{\prime}(\alpha)=0$. Assuming 
$\mathcal{A}(\alpha) \neq 0$ and $\mathcal{A}^{\prime}(\alpha) = 0$ we find $c=1$ while assuming $\mathcal{A}(\alpha)=0$ and $\mathcal{A}^{\prime}(\alpha) \neq 0$ we get $c=-1$. Thus for $Q=0$ the wave function $\mathcal{A}(x)$ becomes an eigenfunction of the global parity operator 
$\hat{O}_F \equiv \hat{\varPi}_{\alpha}$ which performs mirror reflection around the axis located at $\alpha$. Note that, for
$Q=0$, Eq.~(\ref{QJ}) becomes:
$$\left| \widetilde{Q} \right|^{2}=-J^{2},$$
which, in turn implies that $\widetilde{Q}=J=0$. Therefore, for parity, the global symmetric scenario is realized either in bound state problems where the asymptotic conditions are symmetric and $J=0$ or in scattering problems if incoming waves enter the potential $U(x)$ from  both sides in a certain,  symmetric manner so that $J=0$. The latter are actually the \textit{zero-current states} discussed in \cite{Kalozoumis2013a}, denoting that the symmetry is restored in the whole space. As it was shown there, under appropriate asymptotic conditions, the latter can hold even if $U(x)$ is locally symmetric.

\noindent \textit{Translation} ($\sigma=1$): For a global translational symmetry, where $\sigma=1$, we set $Q=0$ in Eq.~(\ref{eigf}) which becomes an eigenvalue equation. Then, from Eq.~(\ref{QJ}) it follows that:
\begin{equation}
\label{module_Q_tilde2} \left| \frac{\widetilde{Q}}{J} \right| =1
\end{equation}
and consequently $\frac{\widetilde{Q}}{J}$ becomes a phase which is in agreement with the fact that it is an eigenvalue of the translation operator $\hat{T}_L$. 

Global translation symmetry implies infinite periodicity for the potential $U(x)$, i.e. $U(x)=U(x+L)$ for all $x$ in $\mathbb{R}$. Thus, the property $U(x)=U(x + n L)$ with $n~\in~\mathbb{Z}$ applies too, implying that Eq.~(\ref{genblp}) can be written replacing the translation parameter $L$ with $n L$. For global translation symmetry, the condition $Q=0$ must hold for all $n$ in 
$\mathbb{Z}$. However the $\widetilde{Q}$s would in general differ for different $n$ values. It is therefore useful to introduce an index denoting as $\widetilde{Q}_{nL}$ the constant $\widetilde{Q}$ corresponding to the displacement $nL$. Equation~(\ref{genblp}) generalizes accordingly:
\begin{eqnarray}
\label{QL}
\mathcal{A}(x+L)=e^{i \theta(L)} \mathcal{A}(x)~~~~~&;&~~~~~\theta(L)=\theta_{\widetilde{Q}_L} \nonumber \\ \widetilde{Q}_L&=&\pm \vert J \vert 
e^{i\theta_{\widetilde{Q}_L}} 
\end{eqnarray}
Then, due to the infinite periodicity we expect that $\widetilde{Q}_L$ will be the same for every $x$ in $\mathbb{R}$. Using Eq.~(\ref{genblp}) we can relate 
$\mathcal{A}(x + n L)$ with $\mathcal{A}(x)$ either by performing $n$ translations by $L$ or one translation by $n L$. This yields
\begin{equation}
\label{phases}
\mathcal{A}(x + n L) = e^{i \theta(n L)} \mathcal{A}(x) = (e^{i \theta(L)})^n
\mathcal{A}(x), 
\end{equation}
implying in turn that
\begin{equation}
\label{phases2}
 \theta(n L) = n \theta(L),
\end{equation}
which means that $\theta(L)=k L$ with $k$ a constant of inverse length dimension. In addition one obtains a relation for the phases of the different 
$\widetilde{Q}$'s:
\begin{equation}
\label{Qtphase}
\theta_{\widetilde{Q}_{nL}}= n \theta_{\widetilde{Q}_L}.
\end{equation}
Equation~(\ref{phases}) can be written as:
\begin{equation}
\label{bloch_periodic} \mathcal{A}(x + nL)=e^{i k nL} \mathcal{A}(x),
\end{equation}
revealing that $k$ is the familiar crystal wave vector. Indeed, multiplying both sides of this relation with $e^{-ik(x+nL)}$ leads to
\begin{equation}
\label{bloch_version_1}
e^{-i kx} \mathcal{A}(x)=e^{-i k(x+nL)} \mathcal{A}(x+nL),
\end{equation}
which is exactly the periodic function $u(x)=u(x+nL)$,  with period $L$, which appears in the Bloch theorem
\begin{equation}
\label{bloch_version_2}
\mathcal{A}(x)=e^{i k x} u(x)~~;~~ u(x)=u(x+nL).
\end{equation} 
Thus it is shown that Eq.~(\ref{genblp}) contains the parity and Bloch theorems as special limits in case of global symmetry restoration.

\subsection{Locally symmetric potentials}

\noindent We turn now to the second intriguing scenario of symmetry breaking. Contrary to the above discussed case where the symmetry of the potential $U(x)=U(\bar{x})$ holds globally i.e. in all of space, this is obviously not the case for locally symmetric potentials. Here, instead of using the corresponding symmetry operators, as was done in Ref.~\cite{Kalozoumis2013a}, we will analyse the local symmetry properties by employing the invariants $Q$ and $~\widetilde{Q}$, which encode directly the effect of the symmetry operation on the wave function (see Eq.~(\ref{genblp})).

In the extreme case of complete breaking of the global symmetry, i.e. when there is no domain $\mathcal{D}_{i}$ for which a remnant of the global symmetry is present in $U(x)$, then, as expected, there is also no domain where $Q$ and $\widetilde{Q}$ are constant. Although, one can still define the spatially dependent function $Q(x),~\widetilde{Q}(x)$, their non-constancy brings no advantage to the representation of the wave mechanical problem.

\begin{figure}[t!]
\centering
\includegraphics[width=.85\columnwidth]{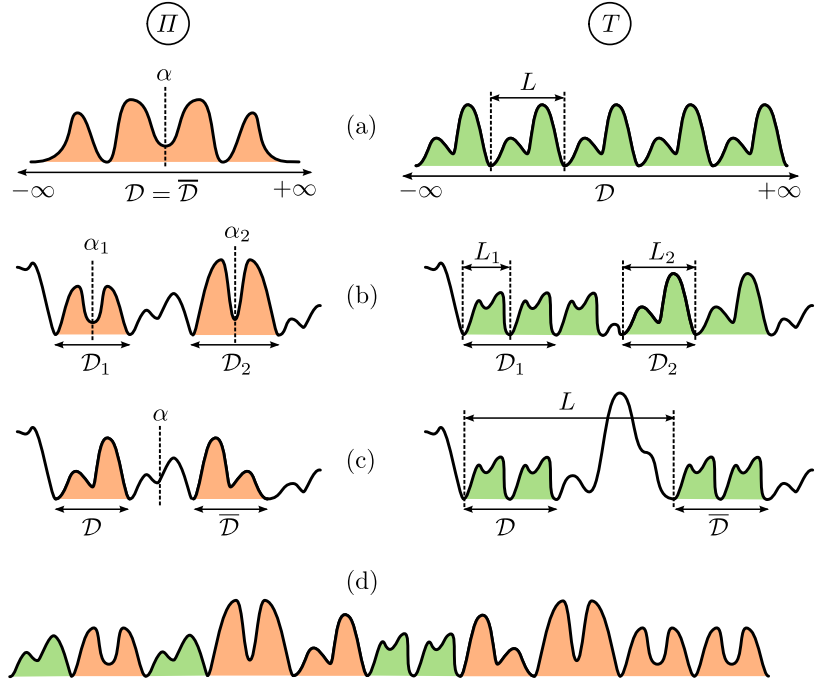}
\caption{
(Color online)
Illustration of different types of global and local symmetries, distinguishing between reflection ($\varPi$) through $\alpha$ or translation ($T$) by $L$. Each symmetry maps a domain $\mathcal{D}$ to $\bar{\mathcal{D}}$: 
(a) global symmetry, (b) nongapped local symmetry, (c) gapped local symmetry, (d) complete local symmetry.   }
\label{fig2}
\end{figure}

Nevertheless, when there are one or several domains $\mathcal{D}_n$ ($n=1,2,..,N$)  where the corresponding symmetry is retained ($U_{n}(x)=U_{n}(\bar{x})~ \forall~x~\in \mathcal{D}_{n}$), then the global symmetry is partially broken and the previous analysis is applicable leading to the existence of the corresponding pair of complex spatially constant quantities $(Q_n,\widetilde{Q}_n)$ in each domain $\mathcal{D}_n$, allowing the determination of the image $\mathcal{A}(\bar{x})$ from $\mathcal{A}(x)$ for all $x$ in $\mathcal{D}_n$ via Eq.~(\ref{genblp}). In addition the relation of these constants to the globally conserved current $J$ gives a constraint between their magnitudes in different domains:
\begin{equation}
\label{conQ}
\vert \widetilde{Q}_1 \vert^2 - \vert Q_1 \vert^2 = \vert \widetilde{Q}_2 \vert^2 - \vert Q_2 \vert^2 = \ldots = \vert \widetilde{Q}_N \vert^2 - \vert Q_N \vert^2.
\end{equation}
Therefore, if local symmetry is present in $U(x)$ one can distinguish the following cases: 

(i)\textit{ Nongapped local symmetries}:
This class of symmetries occurs when a symmetry domain $\mathcal{D}$ coincides, overlaps, or connects with its image $F(\mathcal{D})$. This case is illustrated in Fig.~\ref{fig2}(b) indicating that along the total potential landscape $U(x)$,
one or many such symmetry domains can exist, possibly with nonsymmetric parts in between.
For local reflection ($\varPi$) symmetry, the wave field in one half of ${\mathcal{D}}$ is determined from the field in the other half through the corresponding invariant currents $Q$ and $\widetilde{Q}$, which can be evaluated at the position of the axis of reflection $\alpha_n$. On the other hand, for local translation ($T$) symmetry, the field within the first interval of length $L$ in $\mathcal{D}$ successively determines its images in the translated parts of $U(x)$. Then the pair of invariants $Q,~\widetilde{Q}$ can be evaluated at the corresponding boundaries.

(ii) \textit{Gapped local symmetries}:
In this case, a gap separates the symmetry related domains in configuration space. Within this gap the potential $U(x)$ can obey another or no symmetry, as indicated in Fig.~\ref{fig2}(c). For a connected $\mathcal{D}$, a gapped (local) reflection symmetry occurs if the symmetry axis lies outside the associated domain ($\alpha \notin \mathcal{D}$) and $\mathcal{D}$ is consequently mapped onto $\bar{\mathcal{D}}$ with $\mathcal{D} \cap \bar{\mathcal{D}}=\emptyset$. A gapped $T$ symmetry occurs if the translation length $L$ exceeds the size of $\mathcal{D}$. 
An appealing property emerges if $Q$ and $\widetilde{Q}$ can be evaluated from a pair of symmetry-connected points. Then, Eq.~(\ref{genblp}) maps the wave function from one part of the potential to a remote (symmetry related) part, although there is an arbitrary potential (and thereby field variation) in the intervening gap. 

(iii)\textit{ Complete local symmetry (CLS)}: 
Among the different setups which support the (partial) breaking of a symmetry, the case of systems which are completely decomposable to domains of local symmetry (Fig.~\ref{fig2}(d)) is particularly appealing.  By complete decomposability we refer to the system's property of being decomposable into domains $\mathcal{D}_n$ where the corresponding symmetry is fulfilled
\begin{equation}
\label{potinv1}
U_{n}(x)=U_{n}(\bar{x})~~~~~\forall~~x \in \mathcal{D}_{n},
\end{equation}
$U_{n}(x)$ being the part of $U(x)$ in $\mathcal{D}_{n}$, where the different $\mathcal{D}_{n}$ cover the entire setup.
We refer to such a potential as a \textit{completely locally symmetric (CLS)} potential. This can be realized, in particular, if domains of nongapped local $\varPi$- or $T$ symmetry are attached. The domains can be characterized by a single kind of symmetry or can be of mixed type. Gapped local symmetries can also exist in CLS potentials. In this case, the gaps between their source ($\mathcal{D}_n$) and image ($\overline{\mathcal{D}}_n$) domains are all filled in, either by nongapped local symmetry domains or by the source or image domain of other gapped symmetries.
In the latter case, the possible symmetry domains can be multiply intertwined, rendering the presence of a local symmetry structure far from evident. 
Then the pair of invariants $Q$ and $\widetilde{Q}$ can be utilized as a detection tool for possible  local symmetries, since by calculating them for every pair $(x,\bar{x})$ and using different $\alpha$ or $L$, their constancy would reveal underlying symmetry domains, if present.

\begin{figure}[t!]
\centering
\includegraphics[width=.85\columnwidth]{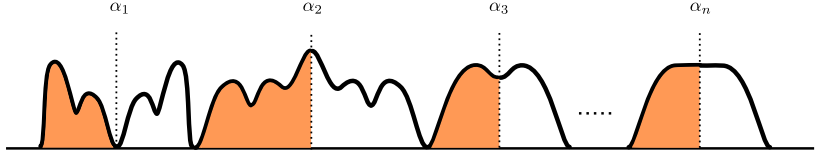}
\caption{\label{fig3}
(Color online) Schematic of a completely locally symmetric setup, with each potential unit being reflection symmetric. The coloured parts of the barriers indicate the part of the potential where we need to know the field $\mathcal{A}(x)$ in order to determine $\mathcal{A}(\bar{x})$, if the invariants $\{Q_{n},\widetilde{Q}_{n}\}$ for each symmetry domain are known.}
\end{figure}

To elaborate more on the class of CLS materials, we consider a system which is comprised of $N$ nonoverlapping domains each one characterized by a different reflection symmetry and which cover entirely the spatial extent of the device.
Assume that in each domain the axis of mirror symmetry lies at the center of the domain. As already discussed the domain is mapped to itself in this case. Using Eq.~(\ref{genblp}) in each domain one can obtain the wave function in the right part of a domain from the wave function of the left part (or the opposite), by employing the invariants  $Q_{n},~\widetilde{Q}_{n}$ of the corresponding potential $U_{n}(x)$ in the subdomain $\mathcal{D}_{n}$. Thus we need only to know the wave function in the half space in order to obtain it in the other half space, as shown in Fig~\ref{fig3}. Therefore, for the $N$ reflection symmetric potential units in each symmetry domain $\mathcal{D}_{n}$, there will exist $N$ couples of $\{Q_{n},~\widetilde{Q}_{n}\}$.  In the case of the global symmetry restoration  the situation is similar, though simplified since the rhs of the field can be calculated from the lhs by a multiplication with $\pm 1$. What is different in the case of a broken local symmetry is the fact that the half-intervals where we need to know the wave function are disconnected forming an array instead of a simply connected region. The aforementioned analysis suggests that the characteristic class of CLS materials generalizes the notion of aperiodic and quasiperiodic systems.

\begin{figure}
\centering
\includegraphics[width=.85\columnwidth]{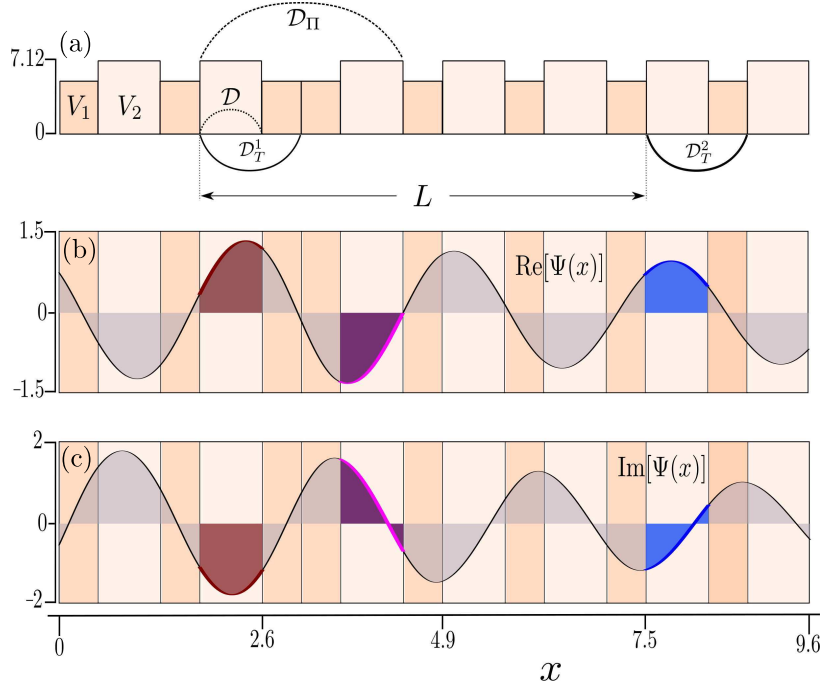}
\caption{ \label{fig4}
(color online)(a) Quantum mechanical setup, comprised of $15$ rectangular barriers with strengths and widths $V_{1}=5.45,~d_{1}=0.5$ and $V_{2}=7.12,~d_{21}=0.8$. (b),~(c) Real and imaginary  parts of the wave function calculated for the energy $\epsilon=12$. Provided that the pair of invariants \{$Q_{\varPi},~\widetilde{Q}_{\varPi}$\}, belonging to the locally reflection symmetric part of the potential defined on $\mathcal{D}_{\varPi}$, are known, then the wave function $\varPsi(x)$ on $\mathcal{D}$ (brown area) can be mapped onto its symmetry related image $\varPsi(\bar{x})$, shown in purple. Accordingly, domains $\mathcal{D}_{T}^{1},~\mathcal{D}_{T}^{2}$ are related via a translational symmetry by a length $L$. The corresponding pair of invariants \{$Q_{T},~\widetilde{Q}_{T}$\} provides the wave function in the symmetry related domain $\mathcal{D}_{T}^{2}$, shown in blue.}
\end{figure}

Figure~\ref{fig4} illustrates a CLS quantum mechanical setup comprised of rectangular barriers of strengths $V_{1},~V_{2}$, which can be decomposed into locally symmetric units in multiple ways. Here we select the local reflection symmetric domain $\mathcal{D}_{\varPi}$ (depicted with the dashed arc) and also the domains $\mathcal{D}_{T}^{1},~\mathcal{D}_{T}^{2}$ (indicated by the solid arcs) which are related via a (local) translational symmetry transformation. Knowing the corresponding pairs of invariants \{$Q_{\varPi},~\widetilde{Q}_{\varPi}$\} and  \{$Q_{T},~\widetilde{Q}_{T}$\} and the wave function $\varPsi(x)$   $\forall~x \in \mathcal{D}$ ($\mathcal{D}$ is the domain of the fourth barrier) and using Eq.~(\ref{genblp}) we can obtain the corresponding images of the wave function in the symmetry related domains. The image under reflection of the wave function in the domain $\mathcal{D}_{\varPi}$, computed from Eq.~(\ref{genblp}), is indicated in the purple colored area (seventh barrier). Accordingly, for the translational symmetry, the image of the wave function in the domain $\mathcal{D}$  
is illustrated in the blue colored area (thirteenth barrier). Note that if one knows the wave function in $\mathcal{D}$ and the corresponding \{$Q_{\varPi},~\widetilde{Q}_{\varPi}$\},  \{$Q_{T},~\widetilde{Q}_{T}$\} pairs, then the mapping is valid independently of the intervening region between the symmetry related domains.

\section{Formulation of scattering via the invariants $Q$ and $\widetilde{Q}$} \label{scattering}

As we have seen, the invariant nonlocal currents $Q,~\widetilde{Q}$ provide a link between space points which are related via a discrete symmetry, namely reflection or translation, providing information for the incoming and outgoing waves on either side of a symmetry domain. From this viewpoint it is of interest to examine whether and how it is possible to express the transfer matrix (TM) in terms of $Q$ and $\widetilde{Q}$.

Let us consider an arbitrary, reflection symmetric scatterer, part of larger setup. To simplify the notation we will omit indexing in domain specific quantities like symmetry axes, invariant currents etc. We assume that from either side of the scatterer the incoming and reflected waves are plane waves, such that either the lhs and the rhs are potential free regions, or the respective potential units support plane waves e.g rectangular barriers. Thus, on either side of the scatterer the incident plane waves are:
\begin{equation} 
\label{plane_wave_1} \varPsi_{L}(x)=A e^{ikx}+B e^{-ikx}
\end{equation}
\begin{equation}
\label{plane_wave_2} \varPsi_{R}(x)=C e^{ikx}+D e^{-ikx}
\end{equation}
The corresponding TM which describes the propagation from the lhs to the rhs is:
\begin{equation}
\label{2d_TM} 
\left(\begin{array}{c}
A \\ B
\end{array} \right) =
\left(\begin{array}{cc}
w & z \\ z^{*} & w^{*}
\end{array} \right)
\left(\begin{array}{c}
C \\ D
\end{array} \right)
\end{equation}
Solving Eq.~(\ref{2d_TM}) with respect to $w,~w^{*}$, respectively, we get:
\begin{equation}
\label{TM_eqs_omega} w=\frac{A-zD}{C}~~~;~~~w^{*}=\frac{B-z^{*}C}{D}.
\end{equation}
Having considered the complex conjugate of the latter and equating the two expressions for $\omega$, we obtain:
\begin{equation}
\label{TM_z_1} z\left(|C|^{2}-|D|^{2}\right)=B^{*}C-AD^{*},
\end{equation} 
which, combined with the expression for the probability current of a plane wave $J=k\left(|C|^{2}-|D|^{2}\right)$, relates the matrix element $z$ with the plane wave coefficients:
\begin{equation}
\label{TM_z_2} z=\frac{\left(B^{*}C-D^{*}A\right)k}{J}.
\end{equation} 
Similarly, the substitution of $w=\frac{A-zD}{C}$ into Eq.~(\ref{TM_z_2}) leads to:
\begin{equation}
\label{TM_omega} w=\frac{\left(C^{*}A-B^{*}D\right)k}{J}.
\end{equation}
We consider now that the scatterer on the domain $\mathcal{D}$ possesses a reflection symmetry and is part of a larger CLS potential. 
On either side of $\mathcal{D}$ and on its boundaries, the wave functions are plane waves given by Eqs.~(\ref{plane_wave_1}),~(\ref{plane_wave_2}). The substitution of Eqs.~(\ref{plane_wave_1}),~(\ref{plane_wave_2}) into Eqs.~(\ref{Q}),~(\ref{Qtilde}), leads to the following expression for the invariants $Q$ and $\widetilde{Q}$:
\begin{equation}
\label{Q_plane_wave} Q=k\left(ACe^{2ik\alpha}-BDe^{-2ik\alpha}\right),
\end{equation} 

\begin{equation}
\label{Q_tilde_plane_wave} \widetilde{Q}=k\left(AD^{*}e^{2ik\alpha}-BC^{*}e^{-2ik\alpha}\right)
\end{equation}
where $\alpha$ denotes the center of the domain $\mathcal{D}$ where the scatterer is positioned.
The solution of Eqs.~(\ref{Q_plane_wave}),~(\ref{Q_tilde_plane_wave}) with respect to $A$ and $B$ yields
\begin{equation}
\label{coef_A} A=-\frac{e^{-2ik\alpha}\left(\widetilde{Q}D-QC^{*}\right)}{J}
\end{equation}
\begin{equation}
\label{coef_B} B=-\frac{e^{2ik\alpha}\left(\widetilde{Q}C-QD^{*}\right)}{J}
\end{equation}
and their substitution into Eqs.~(\ref{TM_z_2}),~(\ref{TM_omega}) leads to:
\begin{equation}
\label{TM_z_ver2} z=-\frac{ke^{-2ik\alpha}}{J^{2}}\left[QC^{*}D^{*}-Q^{*}CD+\widetilde{Q}\left(|C|^{2}+|D|^{2}\right)\right]
\end{equation}

\begin{equation}
\label{TM_omega_ver2} w=-\frac{ke^{-2ik\alpha}}{J^{2}}\left[2\widetilde{Q}C^{*}D-Q(C^{*})^{2}-Q^{*}D^{2}\right].
\end{equation} 
For convenience we define 
\begin{equation}
\label{z_magnitude} \mathcal{W}= \frac{k}{J^{2}}\left[QC^{*}D^{*}-Q^{*}CD+\widetilde{Q}\left(|C|^{2}+|D|^{2}\right)\right],
\end{equation}
where $\mathcal{W}$ is imaginary, since for reflections $\widetilde{Q}$ is imaginary (see Eq.~(\ref{Qtilde_via_real_q}), with $q_{2}=q_{3}$).
This allows to write the transfer matrix element $z$ as
\begin{equation}
\label{z_polar} z=\vert \widetilde{\mathcal{W}} \vert e^{-i(2k\alpha-\frac{\pi}{2})}e^{i\pi(\frac{1-\textrm{sign}(\widetilde{\mathcal{W}})}{2})},
\end{equation}
with $\widetilde{\mathcal{W}}=i\mathcal{W}~\in \mathbb{R}$.
If the center $\alpha$ of the domain $\mathcal{D}$ is at zero, then $z$ is imaginary, as one would expect.
This result is consistent with the fixed phase value $\varphi_{z}=\frac{\pi}{2}$ of the matrix element $z$, in the case of a globally symmetric potential with $\alpha=0$. Nevertheless, here it is shown that in the general case of a potential possessing a reflection symmetric part which covers the domain $\mathcal{D}$, the corresponding TM element $z$ has also a fixed phase, equal to:
\begin{equation}
\label{z_phase} \varphi_z=\frac{\pi}{2}-2k\alpha.
\end{equation}

In order to relate the TM corresponding to the aforementioned reflection symmetric part of the potential to the respective invariants $Q,\widetilde{Q}$, a direct correspondence between $z,~w$ and $Q,\widetilde{Q}$  should be derived. Equations~(\ref{TM_z_ver2}),~(\ref{TM_omega_ver2}) involve the coefficients $C,~C^{*},~D,~D^{*}$, rendering this direct correspondence impossible. To overcome this difficulty one can consider the reflection symmetric potential on the domain $\mathcal{D}$ to constitute the complete scattering potential of the setup, assuming that the potential outside of $\mathcal{D}$ vanishes. Then the reflection symmetry becomes global and the currents $Q,~\widetilde{Q}$ obtain new values $Q_{g},~\widetilde{Q}_{g}$ (index $g$ is for global). In this case the following correspondence for the plane wave coefficients holds:
$$A \rightarrow 1,~ B \rightarrow r,~C \rightarrow t,~D \rightarrow 0,$$
where $1$ is the amplitude of the incident wave and $r$, $t$ the reflection and the transmission amplitudes respectively. Under this assumption, Eq.~(\ref{TM_z_ver2}) becomes:
\begin{equation}
\label{TM_z_ver3} z=-\frac{ke^{-2ik\alpha}}{J_{g}^{2}}\widetilde{Q}_{g} T
\end{equation}
where $T=|t|^{2}$ is the transmission coefficient. By substituting $T=\frac{J_{g}}{k}$ we finally obtain:
\begin{equation}
\label{TM_z_ver4} z=-\frac{\widetilde{Q}_{g}}{J_{g}}~e^{-2ik\alpha},
\end{equation}
where the connection between $z$ and $\widetilde{Q}_{g}$ is direct. Similarly, Eqs.~(\ref{TM_omega}),~(\ref{Q_plane_wave}) become
\begin{equation}
\label{TM_omega_ver3} w=\frac{kt^{*}}{J_{g}},
\end{equation}
and
\begin{equation}
\label{Q_plane_wave_ver2} 
t^{*}=\frac{Q_{g}^{*}}{k}e^{2ik\alpha},
\end{equation}
respectively. Finally, we find that $w$ can be expressed via $Q_{g}$ and the position of the symmetry axis of the domain $\mathcal{D}$:
\begin{equation}
\label{TM_omega_ver4} w=\frac{Q_{g}^{*}}{J_{g}}~e^{2ik\alpha}.
\end{equation}

Therefore, we can express the TM of the reflection symmetric potential corresponding to the domain $\mathcal{D}$, in terms of the invariants  $Q_{g}$, $\widetilde{Q}_{g}$, the current $J_{g}$ and the position of the symmetry axis $\alpha$. Needless to say that the TM associated to the domain $\mathcal{D}$ (determined by Eqs.~(\ref{TM_z_ver4}),~(\ref{TM_omega_ver4})) does not depend on the form of the potential outside of $\mathcal{D}$.

\section{Symmetry scales and perfectly transmitting resonances in locally symmetric systems} \label{scales}

\begin{figure}[t!]
\centering
\includegraphics[width=.85\columnwidth]{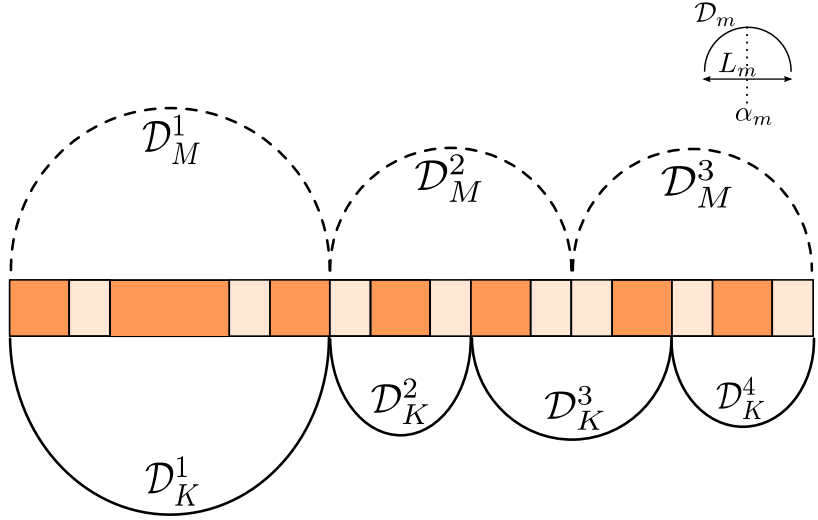}
\caption{\label{fig5} (Color online) Aperiodic, CLS setup. Each color box stands for a different (LP symmetric) scatterer, corresponding to i.e. a quantum or a photonic individual scatterer. The dashed and solid arcs depict $3$ and $4$ LP symmetric units corresponding to two different decompositions $M,~K$ respectively. As an example, a single domain $\mathcal{D}_{m}$ has length $L_{m}$ and extends around the local symmetry axis $\alpha_{m}$.}
\end{figure}

We will focus here on CLS setups where the relevant symmetry transform is the reflection. Following the terminology of Refs.~\cite{Kalozoumis2013a,Kalozoumis2013b}, we will refer to the reflection symmetry operation which is valid into each of the restricted spatial domains of the setup as \textit{local parity} (LP) symmetry. In turn, we will refer to the parts of the potential which possess local reflection as LP symmetric units. The decomposition of a CLS setup into parts with LP symmetry, may occur in more than one way and a setup which offers this possibility is regarded as one with multiple symmetry scales. In Fig.~\ref{fig5} we show an aperiodic CLS material consisting of $N$ scatterers which can be grouped in $M$ LP symmetric, non-overlapping units, covering the domains $\mathcal{D}_{M}$, respectively. Nevertheless, the $N$ scatterers can be also grouped in $K$ different LP, non-overlapping units, corresponding to the domains $\mathcal{D}_{K}$. Each different color box stands for a different LP scatterer. In a quantum system these could correspond to barriers or wells whereas in photonic systems they correspond to dielectrics of different refraction index. The arcs delimit the domains where the LP symmetry is valid and which comprise each decomposition. Particularly, the dashed and solid arcs denote two possible--but not the only--different decompositions ($M,~K$) in LP symmetric domains, consisting of $3$ and $4$ LP symmetric units ($\mathcal{D}_{M},~\mathcal{D}_{K}$).
Note, that in each decomposition the domains which constitute it, should  be non-overlapping. 

In Refs.~\cite{Kalozoumis2013a,Kalozoumis2013b} it has been shown that the multitude of the possible decompositions of an aperiodic or quasiperiodic, completely LP symmetric device is strongly related to its transmission properties and specifically to the emergence of perfect transmission resonances (PTRs). 

\subsection{Sum rules emerging from local invariants}

Let us return to the definition of the first nonlocal invariant quantity $Q$, which was introduced in Section~\ref{invariants}. Here we restrict ourselves to reflection transforms 
\begin{equation}
\label{parity_trans} F(x)=2\alpha_{m}-x,
\end{equation} 
where $\alpha_{m}$ is the position of the symmetry axis of the corresponding LP symmetric unit.
For a such a CLS setup which is decomposable in $N$ LP symmetric units, there exist $N$ generally different pairs of invariants $Q_{m},~\widetilde{Q}_{m}$ ($m=1,...N$), each one corresponding to the domain $\mathcal{D}_m = [x_{m-1}, x_m]$.
Note here the value of each $Q_m,~\widetilde{Q}_{m}$ depends on the considered local symmetry decomposition and on the input (generalized) energy $\epsilon$.

The importance of the invariants $Q$ and $\widetilde{Q}$ was revealed in Section~\ref{Bloch} with the generalization of the Bloch and parity theorems. In~\cite{Kalozoumis2013b} we have used $Q$ to obtain a new quantity--denoted as $\mathcal{L}$--which refers globally to the (CLS) system and enables the classification of perfect transmission resonances according to the corresponding field module configurations in terms of local (parity) symmetries. Particularly, we have found that PTRs can be classified in the following two categories:

\noindent \textit{Asymmetric} PTR ($a$-PTR) is a
$T = 1$ stationary wave whose field magnitude
$u(x)$, in general, does \textit{not} follow the reflection symmetries of the potential units which comprise the corresponding decomposition of the CLS potential landscape. 

\noindent \textit{Symmetric} PTR ($s$-PTR) is a resonance which resonates with $T_m = 1$ in each subdomain $D_m$ of a considered decomposition in reflection symmetric potential units. Here $T_m$ is the transmission coefficient through $D_m$ alone. The wave field magnitude $u(x)$ follows is completely locally symmetric in the sense that it follows the reflection 
symmetry of each unit which constitutes the corresponding decomposition (as was shown in Refs.~\cite{Kalozoumis2013a}). In this PTR case the wave field magnitude is $u(x) = 1$ at both boundaries of any subdomain $D_m = [x_{m−1} ,x_m ]$.

In both cases, depending on whether the setup is decomposed in an even or odd number of reflection symmetric units, the value of $\mathcal{L}$ becomes $0$ or $\kappa$, respectively.

By using Eq.~(\ref{genblp}) we will show that $\mathcal{L}$ expresses, in fact, a sum rule for the pairs of $Q_{m},\widetilde{Q}_{m}$ characterizing the corresponding LP symmetric domains. Furthermore, we will examine the energy dependence of  $\vert \mathcal{L} \vert$ and particularly its relation to the transmission spectrum of a CLS system.  

The extended derivation of $\mathcal{L}$ can be found in Ref.~\cite{Kalozoumis2013b}. Here we give the corresponding expression: 
\begin{equation}
\mathcal{L}= \sum_{m=1}^{N} (-1)^{m-1} \mathcal{V}_m =\frac{1}{2i}\left(\frac{\mathcal{A}'(x_0)}{\mathcal{A}(x_0)} - (-1)^{N}~ \frac{\mathcal{A}'(x_N)}{\mathcal{A}(x_N)}\right) 
\label{conserved_q_divided_sum} 
\end{equation}
where
\begin{equation}
\label{chap6_Q_divided}
\mathcal{V}_m \equiv \frac{Q_m}{\mathcal{A}(x_{m-1})~\mathcal{A}(x_{m})}.
\end{equation}
Equation~(\ref{conserved_q_divided_sum}) clearly indicates that $\mathcal{L}$ depends only on the field at the global boundaries $x_{0},~x_{N}$ of the setup. 
The replacement of $\mathcal{A}(x)$ by the map provided in Eq.~(\ref{genblp}), yields
\begin{equation}
\label{chap6_Q_divided_new}
\mathcal{V}_m = \frac{Q_m}{\mathcal{A}(x_{m-1})~\left[\frac{\widetilde{Q}_{m}}{J}\mathcal{A}(x_{m-1})-\frac{Q_{m}}{J}\mathcal{A}^{*}(x_{m-1})\right]}.
\end{equation}
In the same manner, we replace $\mathcal{A}(x_{m-1})$ and finally obtain a recursive relation which only includes the $Q,~\widetilde{Q}$ values corresponding to each symmetric subdomain and the field value at the device's starting point $\mathcal{A}(x_{0})$. Since this property is valid for every $\mathcal{V}_{m}$ assigned to each symmetric subdomain, it will also hold for the $\mathcal{L}$. 

On the other hand, $\mathcal{L}$ is a global quantity, which characterizes the device as a whole. Assuming asymptotic conditions given by
\begin{equation} 
\label{incident} \mathcal{A}_{x<x_{0}}(x)=e^{i\kappa x}+re^{-i\kappa x}
\end{equation}
\begin{equation}
\label{transmitted} \mathcal{A}_{x>x_{N}}(x)=te^{i\kappa x},
\end{equation}
on either side of the device  (computed at $x=x_{0}$ and $x=x_{N}$, respectively) and  substituting them in Eq.~(\ref{conserved_q_divided_sum}) we get the following expressions for $\mathcal{L}$ ($r$ and  $t$ are the reflection and transmission amplitudes)
\begin{equation}
\label{chap6_L_even} \mathcal{L} =-\frac{\kappa r}{1+r}~~~;~~~ N~\text{even} 
\end{equation} 
and
\begin{equation}
\label{chap6_L_odd} \mathcal{L} =\frac{\kappa}{1+r}~~~;~~~ N~\text{odd},
\end{equation}  
for an even and odd number of LP symmetric units, respectively. In the case of a PTR ($r=0$) $\mathcal{L}$ vanishes for even $N$ or, for odd $N$, becomes $\mathcal{L}=\kappa$, which is in accordance with the PTR classification presented in~\cite{Kalozoumis2013b}. Equation~(\ref{chap6_L_even}) clearly indicates that if $r=0$ (PTR) then $\mathcal{L}=0$ and inversely, if $\mathcal{L}=0$ then $r=0$.
On the other hand, Eq.~(\ref{chap6_L_odd}) when $r=0$ becomes $\mathcal{L}=\kappa$. Inversely, if $\mathcal{L}=\kappa$, then
\begin{equation}
\label{inverse_L_odd} \kappa=\frac{\kappa}{1+r} \Rightarrow r=0
\end{equation}
Finally, the  separation of Eqs.~(\ref{chap6_L_odd}),~(\ref{chap6_L_even}) into real and imaginary parts yields ($r=r_{R}+ir_{I}$)
\begin{equation}
\label{chap6_L_even_real_imag} \mathcal{L} =\frac{2 \kappa r_{I}}{\left[u(x_{0})\right]^{2}}-i\left[\frac{2 \kappa r_{R}+2 \kappa R}{\left[u(x_{0})\right]^{2}} \right]~~~;~~~ N:~\text{even} 
\end{equation} 
and
\begin{equation}
\label{chap6_L_odd_real_imag} \mathcal{L} =\frac{2 \kappa r_{I}}{\left[u(x_{0})\right]^{2}}+i\left[\frac{2 \kappa +2 \kappa R}{\left[u(x_{0})\right]^{2}} \right]~~~;~~~ N:~\text{odd},
\end{equation}   
which shows that the expressions for $\mathcal{L}$ for a decomposition with an even or odd number of LP symmetric units, retain their real part and  differ only with respect to the imaginary. Note that $R=|r|^{2}$ is the reflection coefficient.

Finally, we come back to the treatment of the wave mechanical problem via the invariant components ($q_{1},~q_{2},~q_{3},~q_{4}$) of the invariants $Q,~\widetilde{Q}$ (see Sec.~\ref{Bloch}) and we will discuss in the remaining part of this section some of the relevant specific cases. Using Eqs.~(\ref{magn_map}),~(\ref{phase_map}) we focus on the following two interesting cases emerging for either globally or locally reflection ($\varPi$) symmetric potentials. In this case $q_{2}=q_{3}$ which can be shown by using the mirror axis point in Eqs.~(\ref{q2_invariant}),~(\ref{q3_invariant}).

(i) Consider first a CLS device decomposable into reflection symmetric potential parts. If this system exhibits an $s$-PTR at some energy $\epsilon$, then within each symmetry domain $\mathcal{D}_{m}$ (corresponding to each reflection symmetric potential part) the respective transmission coefficient is $T_{m}=1$ and the wave field magnitude is reflection symmetric. Accordingly, the TM element $z$ of each reflection symmetric potential part is zero, in order that the condition $T_{m}=1$ is fulfilled. In such a case Eq.~(\ref{TM_z_ver4}) implies that $\widetilde{Q}_{m}=0$ for every symmetry domain $\mathcal{D}_{m}$ leading to the condition $q_{1}+q_{4}=0$. Additionally, $Q=2q_{1}+2iq_{2}$, implying that $\tan[\vartheta_{Q}]=q_{1}/q_{2}$, where $\vartheta_{Q}$ is the phase of $Q$. Based on this, Eq.~(\ref{phase_map}) becomes
\begin{equation}
\label{phase_map_2} \tan[\varphi(\bar{x})]=\tan[\vartheta_{Q}-\varphi(x)].
\end{equation} 
Finally, we obtain the following relation for $\varphi(x)$
\begin{equation}
\label{phase_map_3} \varphi(\bar{x})=\vartheta_{Q}-\varphi(x)+n\pi~~~;~~~n=0,~\pm 1,~\pm 2...
\end{equation} 
which indicates that the phase of the wave field is always antisymmetric with respect to the point $\vartheta_{Q}+n\pi$.

(ii) The special second case where $q_{1}=q_{4}=0$, indicates that $\operatorname{Re}[\mathcal{A}(x)]$ and $\operatorname{Im}[\mathcal{A}(x)]$ are symmetric (or antisymmetric) within the reflection symmetric potential unit. Numerical evidence confirms that this case corresponds to an $s$-PTR for a globally reflection symmetric potential.  Note that for incoming waves only from one side of the setup, it is not possible that both $\operatorname{Re}[\mathcal{A}(x)]$ and $\operatorname{Im}[\mathcal{A}(x)]$ are symmetric or antisymmetric. Therefore, this case corresponds to symmetric $\operatorname{Re}[\mathcal{A}(x)]$ and antisymmetric $\operatorname{Im}[\mathcal{A}(x)]$ or vice-versa.  Then, Eqs.~(\ref{magn_map}),~(\ref{phase_map}) combined with Eq.~(\ref{cur_q}) imply that on either side off the mirror symmetry axis the magnitude and phase are symmetric and antisymmetric, respectively.
 
\subsection{Energy dependence of $\mathcal{L}$ and PTRs in an aperiodic quantum system} \label{energy_dependence}

\begin{figure}[t!]
\centering
\includegraphics*[width=.85\columnwidth]{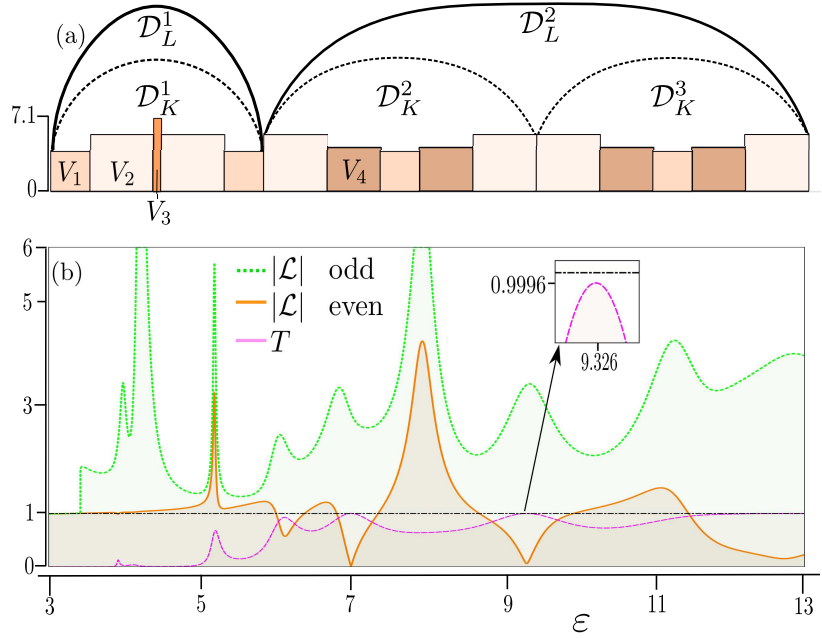}
\caption{\label{fig6} (Color online) (a) Schematic of a quantum mechanical setup comprised of four different types of barriers with strengths $V_{1}=3.4197,~V_{2}=5,~V_{3}=7.0806,~V_{4}=3.0671$ and widths $d_{1}=0.5,~d_{2}=0.8,~d_{3}=0.1,~d_{4}=0.6751$. The dashed arks indicate the decomposition $K$ consisting of three LP symmetric units, and the solid arks depict the two LP symmetric domains of decomposition $L$. 
(b) The green, dotted curve shows $|\mathcal{L}(\epsilon)|$ for the (odd) decomposition $K$ shown in (a). The orange, solid curve illustrates $|\mathcal{L}(\epsilon)|$ for the (even) decomposition $L$. The magenta, dashed line corresponds to the transmission coefficient $T(\varepsilon)$. Note how a very small deviation from $T=1$ (as shown in the inset), is magnified under the prism of $|\mathcal{L}(\epsilon)|$.} 
\end{figure}

We investigate now the energy dependence of $\vert \mathcal{L} \vert$ and the possible implications this might have on the perfect transmission properties of the system.  To this aim, we consider a quantum mechanical, aperiodic setup comprised of four different kinds of rectangular barriers, $V_{1},~V_{2},~V_{3},~V_{4}$. The device belongs to the class CLS of materials, since it can be completely decomposed into LP symmetric potential units. In Fig.~\ref{fig6} two possible such decompositions are shown. The first ($K$), depicted by the dashed arcs, consists of an odd number of LP symmetric units, corresponding to the domains $(\mathcal{D}^{1}_{K},~\mathcal{D}^{2}_{K},~\mathcal{D}^{3}_{K})$. The other one ($L$), indicated by the solid arcs, is comprised of an even number of LP symmetric units corresponding to the domains $\mathcal{D}^{1}_{L},~\mathcal{D}^{2}_{L}$.   
 
Figure~\ref{fig6}(b) illustrates the energy dependence of $|\mathcal{L}|$ for both (odd and even) decompositions. In order to investigate the possible links between $|\mathcal{L}|$ and the transmission properties of the system, the transmission coefficient $T$ spectrum is shown too. First, we consider the case of the $K$ decomposition with $N=3$ LP symmetric units. From Eq.~(\ref{chap6_L_odd}) we find the square magnitude of $\mathcal{L}$ as 
\begin{equation}
\label{chap6_L_odd_magnitude}  |\mathcal{L}|^{2}=\frac{k^{2}}{1+R+2\operatorname{Re}[r]},~~~N:~\text{odd}
\end{equation} 
which is illustrated with the green, dotted line, having obviously an increasing overall trend. 

For a decomposition comprised of an even number of LP symmetric units, $|\mathcal{L}|^{2}$ acquires more interesting properties. As it is given by
\begin{equation}
\label{chap6_L_even_magnitude } |\mathcal{L}|^{2}=\frac{k^{2}R}{1+R+2\operatorname{Re}[r]},~~~N:~\text{even},
\end{equation}
easily one ascertains that in the case of a PTR, where for the reflection coefficient holds that $R=0$. Also, the existence of  $R$ in the numerator indicates the competition with the respective terms in the denominator and as a result, $|\mathcal{L}|$ does not show an overall increase. 

In the PTR case ($T=1$) it is obvious that $|\mathcal{L}|=0$. An interesting feature is indicated in the inset of Fig.~\ref{fig6}. The transmission peak near $\epsilon \simeq 9.326$ seems to be perfectly transmitting. Nevertheless, the closeup in the inset reveals that the transmission is not perfect. The resulting small deviation from the perfect transmitting case is significantly magnified in the quantity $|\mathcal{L}(\epsilon)|$ which deviates substantially from zero. Therefore, $|\mathcal{L}(\epsilon)|$ could possibly act as a `magnifying glass' in situations where there is ambiguity on whether a state corresponds to a PTR or not. A similar `magnification' is demonstrated for the close to $T=1$ plateau near $\varepsilon=13$. At the energy $\varepsilon=7$ we encounter a real PTR, which is constructed according to the PTR construction procedure proposed in Refs.~\cite{Kalozoumis2013a,Kalozoumis2013b}. In this case $|\mathcal{L}(\epsilon)|$ is exactly zero.

\section{Conclusions} \label{conclusions}

It has been shown that structures which are globally or, in particular, locally symmetric with respect to reflections or translations, possess pairs of nonlocal invariant currents $Q$ and $\widetilde{Q}$, for every domain of symmetry.
These invariants characterize generic wave propagation within arbitrary symmetry domains, giving the necessary information to map the wave function between two symmetry related domains.
Within this theoretical framework, the parity and Bloch theorems are generalized for globally broken reflection and translation symmetries. The identification of these invariant currents provides a systematic pathway to the breaking of global discrete symmetries.
A nonvanishing $Q$ indicates the breaking of the corresponding global symmetry while a zero $Q$ denotes the restoration of the global symmetry. In this sense, $Q$ can be regarded as a local remnant of the corresponding (broken) global symmetry.
The general wave mechanical framework adopted here enables the implementation of our formalism to a wide range of (classical or matter wave) scattering systems in, e.g., nanoelectronic devices,
photonic multilayers or acoustic waveguides.
It is also suggested that structures consisting exclusively of locally symmetric building blocks define a new class of materials--completely locally symmetric (CLS) materials--which should be of particular significance concerning the control of certain features of their transmission properties.

It has also been shown how the possible local symmetries of a setup may enable an alternative scattering formulation, by expressing the transfer matrix of an arbitrary (symmetry) domain through the corresponding invariant currents $Q,~\widetilde{Q}$. The connection of the local symmetries with the scattering properties of the system has been stressed by focusing on the effect of the coexistence of multiple symmetry scales. Finally, it was demonstrated how an appropriate sum rule of the invariant $Q$ of each symmetry domain leads to the quantity $\mathcal{L}$ which characterizes globally the device. The energy dependence of $\mathcal{L}$ revealed that it could be potentially used for distinguishing perfectly transmitting resonances.

{\it Acknowledgements.}---We thank V. Zampetakis and M. Diakonou for fruitful and illuminating discussions.

\end{document}